%Paper: hep-th/9307146
%From: Masahiro Maeno <maeno@oskth.kek.jp>
%Date: Sat, 24 Jul 93 10:27:33 +0900

\input phyzzx
\font\fontA=cmssi12

\font\fontB=cmss17
\font\fontC=cmmi10 scaled \magstep2
\def\contracter{$\bracelu\leaders\vrule\hfill\braceru$}
\def\contract#1{\mathop{\vtop{\ialign{##\crcr$\hfil\displaystyle{#1}\hfil$
\crcr\noalign{\kern3pt\nointerlineskip}\contracter\crcr\noalign{\kern3pt}}}}
\limits}

\def\expl#1{{\hbox {\fontC e}}^{\displaystyle #1 }}
\def\bcl{{}_\diamond^\diamond}

\def\eg{{\it e.g.}}
\def\ie{{\it i.e.}}
\def\HALF{{1\over2}}

\def\cr{\cropen{10pt}}

\def\Hint{H_{\hbox{int}}}
\def\Hfree{H_{\hbox{free}}}
\Pubnum={OU-HET-175B}
\date={\fontA July 1993}
\titlepage
\pagenumber=1
\title{\fontB $k^+=0$ Modes in Light-Cone Quantization}
\author	{Masahiro~MAENO\footnote{\dagger}{e-mail address:maeno@jpnyitp.bitnet}}
\address{Department of Physics, Osaka University\break
          Toyonaka, Osaka 560, ~JAPAN}
\abstract{
We investigate the light-cone quantization of $\phi^3$ theory in 1+1 dimensions
with a regularization of discretized light-cone momentum $k^+$.
Solving a second-class constraint associated with the $k^+=0$ mode, we show
that the $k^+=0$ mode propagates along the internal lines of Feynman diagrams
in any order of perturbation, hence our theory recovers the Lorentz invariance.
}
\endpage
\REF\LCQA{T.~Maskawa and K.~Yamawaki, Progr. Theor. Phys. {\bf 56}(1976)270.}
\REF\LCQB{M.~Ida, Nuov. Cim. Lett. {\bf 15}(1976)249.}
\REF\LCQC{N.~Nakanishi and K.~Yamawaki, Nucl. Phys. {\bf B122}(1977)15.}
\REF\Burkardt{M.~Burkardt and A.~Langnau, Phys.~Rev. {\bf D44}(1991)1187.}
\REF\VAC{Th. Heinzl, St. Krusche, E. Werner, Z. Phys. {\bf A334}
(1989)443;\break Phys. Lett. {\bf B256}(1991)55.}
\REF\SP{Th. Heinzl, St. Krusche, E. Werner, Phys. Lett. {\bf B272} (1991)54.}
\REF\McCartor{G.~McCartor and D.~G. Robertson, Z. Phys. {\bf C53}(1992)679.}

\chapter{Introduction}
The light-cone quantization is a useful tool in particle physics.
However, it is a long-standing problem how to treat the mode with the vanishing
light-cone momentum $k^+$.
 \refmark{\LCQA-\LCQC}
In conventional calculation, one simply ignores such modes.
However, the absence of the $k^+=0$  modes causes some problems, \eg, the
breakdown of Lorentz invariance, \refmark\Burkardt
uneasiness about the vacuum definition, etc.
The subtlety comes from constraints over the zero momentum mode.
In the light-cone frame, the conjugate momentum of bosonic field does not
contain $x^+$- derivative (where $x^+={1\over \sqrt2}(x^0+ x^1)$),
hence the definition of momentum must be considered as a second-class
constraint. The theory must be quantized by the Dirac bracket, in which the
$k^+=0$ mode generates secondary constraint.
The secondary constraint includes the information of the interaction.

Recently, some researchers surveyed the problem from the point of view of
vacuum (non-)triviality, \refmark\VAC
and the spontaneous break-down of symmetry. \refmark\SP
McCartor and Robertson investigated light-cone quantization of discretized 3+1
dimensional Yukawa theory taking the modes into account. \refmark\McCartor
They solved the zero mode equation, and showed that the modes propagate in
internal lines of Feynman diagrams, and the inclusion of the modes improves bad
behaviors of loop amplitudes.
In their model, it is rather easy to solve the constraint, because the equation
is linear with resptect to the zero mode.

In this letter, we study the light cone problem of $\phi^3$-theory in a 1+1
dimensions.
 Although the constraint equation is not linear, it exhibits some intersting
feature.
In the following, we solve the equation perturbatively, and
show that McCartor and Robertson's result is able to be extended in any order
of perturbation for this case.

This letter is organized as follows:
In section 2, we quantize the 1+1 dimensional $\phi^3$-theory imposing a
periodic boundary condition for $x^-={1\over\sqrt2}(x^0-x^1)$.
In section 3, we solve the zero mode constraint in each order of coupling
constant.
Substituting the solution into the interaction hamiltonian of the model, we
show that the $k^+=0$ mode propagates along all internal lines of Feynman
diagrams.
In section 4, a problem of our normal ordering is discussed.
The last section is devoted to conclusion.

\chapter{$\phi^3$-model in 1+1 dimension}
To make the problem tractable, we consider a $\phi^3$-model in 1+1 dimensions.
The lagrangian density of the model is
$$
\eqalign{
{\cal L } = \partial_+ \phi \partial_- \phi -{ 1 \over 2 }m^2\phi^2 -
{\lambda\over 3!}\phi^3.
}\eqn\Lagrangean
$$

We impose a periodic boundary condition $\phi(x^-=-{ V \over 2 })=\phi(x^-={ V
\over 2 })$, from which the light-cone momentum is discretized as
$p^+={2\pi\over V}n$ ($n$ is an integer).
Splitting the field $\phi$ as $\phi=\tilde\phi+\phi_0$
(where $\phi_0={ 1 \over V }\int dx^- \phi(x^-)$),
the lagrangian can be written as
$$
\eqalign{
L=&\int dx^-\partial_+ \tilde \phi \partial_- \tilde \phi -\int dx^-{ 1 \over 2
}m^2\tilde\phi^2
 -{ 1 \over 2 }m^2V\phi_0^2\cr
& - \int dx^-{\lambda\over
3!}(\tilde\phi^3+3\phi_0\tilde\phi^2)-{\lambda\over3!}V\phi_0^3.
}\eqn\lagrangian
$$
Here, we do not consider the ordering of operators. It will be discussed later.

{}From \lagrangian, the conjugate momenta are defined as
$\tilde \Pi(x) \equiv { \delta L \over \delta \partial_+\tilde\phi(x)
}=\partial_-\tilde\phi(x)$ and $\Pi_0\equiv { \delta L \over \delta
\partial_+\phi_0}=0$.
The momentum $\Pi_0$ generates a secondary constraint:
$$
\eqalign{
-m^2V\phi_0 - { \lambda \over 2 }V\phi_0^2 - { \lambda\over2}\int dx^- \tilde
\phi^2=0.
}\eqn\second
$$
The canonical momenta have no fields with time ($x^+$)- derivative.
Using the Dirac bracket quantization, we obtain canonical commutation relations
as follows.
$$
\eqalign{
[\phi_p,\phi_q]={1\over 2p^+}\delta_{p+q,0},~~~~
[\phi_p,\phi_0]=-{1\over 2p^+}{ \lambda \phi_p \over m^2-\lambda\phi_0},
}\eqn\CCR
$$
where $\phi_p$ is an expansion coefficient of $\phi$ defined by
$$
\eqalign{
\tilde \phi(x^+=0,x^-)={1\over \sqrt V}\sum_{p\ne0}\phi_p \expl{-ip^+x^-}.
}\eqn\coefficient
$$

In the following, we proceed the calculation in the interaction picture.
The hamiltonian associated with lagrangian \lagrangian\ is divided as
$H=\Hfree+\Hint$, where
$$
\eqalign{
\Hfree&={ 1 \over 2 }m^2\int dx^- \tilde \phi^2,\cr
\Hint&={ 1 \over 2 }m^2V\phi_0^2
+{ \lambda \over 3! }V\phi_0^3
+{ \lambda \over 3! }\int dx^- (\tilde\phi^3+3\phi_0\tilde\phi^2).
}\eqn\hint
$$
We include $\HALF m^2V\phi_0^2$ term (zero-mode mass term) not in $\Hfree$ but
in $\Hint$.

Free Heisenberg equations
$$
\eqalign{
i\dot\phi_p=[\Hfree,\phi_p]={m^2\over2p^+}\phi_p,~~~~i\dot\phi_0=
[\Hfree,\phi_0]=0
}\eqn\Heisenberg
$$
are consistent with the free equation of motion
$\partial_+\partial_-\phi-m^2\phi=0$.
They decide $x^+$-dependence of $\phi_p$ in interaction picture.

{}From the commutation relation \CCR\ and the definition of vacuum
$$
\eqalign{
\phi_p|0\big>=0 ~~~~~~~~~~\hbox{( for $p>0$ )},
}\eqn\vaci
$$
the free propagator of $\phi$ can be calculated as follows:
$$
\eqalign{
\tilde\Delta(x,y)=&\left<T\left(\tilde\phi(x^+,x^-)\tilde\phi(y^+,y^-)\right)
\right>\cr
=&\theta(x^+ - y^+){ 1 \over V }\sum_{p^+>0}{ 1 \over 2p^+
}\expl{-i(p^+(x^--y^-)+{m^2\over 2p^+}(x^+-y^+))}\cr
+&\theta(y^+ - x^+){ 1 \over V }\sum_{p^+>0}{ 1 \over 2p^+
}\expl{i(p^+(x^--y^-)+{m^2\over 2p^+}(x^+-y^+))}
}\eqn\prop
$$
In the conventional equal-time quantization where $x^0$ is chosen as the time,
the discretized expression of propagator is expressed as
$$
\eqalign{
\Delta&(x,y)
=\Delta_0(x,y)+\tilde\Delta(x,y)\cr
&={ i \over 2\pi V }\int_{-\infty}^\infty dp^-\sum_{p^=-\infty}^\infty{ 1 \over
2p^+p^- -m^2+i\epsilon }\expl{-i(p^+(x^- - y^-) + p^-(x^+ - y^+))}
}\eqn\proptwo
$$
where $\Delta_0$ is $p^+=0$ part of $\Delta$, \ie,
$$
\eqalign{
\Delta_0(x,y)=&{ i \over 2\pi V }\int_{-\infty}^\infty dp^-{ 1 \over -m^2
}\expl{-ip^-(x^+-y^+)}={ -i \over m^2V }\delta(x^+-y^+).
}\eqn\propzero
$$
The propagator \proptwo\ is covariant in continuum limit
($V\rightarrow\infty$).

If we simply ignore $\phi_0$, the propagator $\big<T(\phi\phi)\big>$ does not
provide $\Delta_0$ part. However, taking the zero-mode constraint \second\ into
account, $\Delta_0$ is recovered at least in all internal lines of Feynman
diagrams.
The purpose of our discussion below is to demonstrate the recovery of
$\Delta_0$.

\chapter{Solving $\phi_0$-constraint}

{}From the constraint \second, $\phi_0$ is given in terms of $\tilde\phi$ as
$$
\eqalign{
\phi_0= -{ m^2 \over \lambda }\left(-1\pm\sqrt{1-{\lambda^2\over m^4V}\int dx
\tilde\phi^2}\right)
}\eqn\phizeroone
$$
Since the negative sign in the parenthis corresponds to a classically unstable
vacuum $\phi_0={2m^2\over \lambda}$, we ignore it and choose the positive sign
solution, which is expressed in a power expansion form as
$$
\eqalign{
\phi_0=-{m^2\over\lambda}\sum_{n=1}^\infty { (2n-3)!! \over n!
}\left({\lambda^2 \over 2m^4V}\int dx \tilde\phi^2\right)^n.
}\eqn\phizerotwo
$$

Substituting \phizerotwo\ into \hint, the interaction hamiltonian $\Hint$ is
expressed as
$$
\eqalign{
\Hint =\sum_{n>0}\Hint^{(n)}+{ \lambda \over 3! }\int dx^- \tilde\phi^3
}\eqn\hintn
$$
where
$$
\eqalign{
\Hint^{(n)}=-{ m^6V \over \lambda^2 }{ (2n-3)!! \over (n+1)! }~~\bcl\left({
\lambda^2 \over 2m^4V }\int dx^-\tilde\phi^2\right)^{n+1}\bcl.
}\eqn\hintntwo
$$
The symbol $\bcl~~~\bcl$ denotes a certain {\it normal ordering}. Note however
that it is not an usual normal ordering, but it is defined in the next section.

{}From \hintn, the exponent of $\expl{-i\int\Hint dt}$ has terms as follows:
$$
\eqalign{
-i \int \Hint^{(n)}dt =
(-i\lambda)^{2n}{ (2n-3)!!  \over (n+1)! }({-i\over
m^2V})^{2n-1}V^{n-1}\bcl\int dt \prod_{i=1}^{n+1}dx^-_i
\HALF\tilde\phi^2(x_i)\bcl.
}\eqn\hintzero
$$
The above expression describes $2n$ cubic interaction vertices which are
connected one another by $2n-1$ $\Delta_0$ propagators.
The factor $(-i\lambda)^{2n}$ implies that $\Hint^{(n)}$ describes $2n$th-order
interactions where the factor ${\displaystyle ({-i\over m^2V})^{2n-1}}$ comes
from $2n-1$ $\Delta_0$ propagators which combine $2n$ vertices.
The operators $(\HALF\tilde\phi^2)^{n+1}$ remain uncontracted.
The rest of vertices, of which number is $n-1=2n-(n+1)$, are fully contracted.
Since $\Delta_0$ is independent of $x^-$, $x^-$-integrations of fully
contracted vertices are trivially worked out, and generate the factor
$V^{n-1}$.

The factor ${ 1 \over (n+1)! }$ will be canceled by a number of combinations
when we contract all $(\HALF\int\tilde\phi^2)^{n+1}$.
The factor $(2n-3)!!$ is a number of patterns to connect $n+1$ points with
$n-1$ lines using cubic interactions without making loop.
It can be proved by induction as follows.

First, for $\Hint^{(1)}$, the statement is true because there is only one way
to connect two points by one line ($(2\times1-3)!!=1$).
Second, we will show that if the number of contractions in $\Hint^{(k)}$ is
$(2k-3)!!$, the number in $\Hint^{(k+1)}$ is $(2k-1)!!$.
A Feynman diagram in $\Hint^{(k+1)}$ is obtained by adding one
$\HALF\tilde\phi^2$ to $\Hint^{(k)}$.
The $\Delta_0$ propagator should be connected from the new $\HALF\tilde\phi^2$
to somewhere  in $\Delta_0$ propagators in $\Hint^{(k)}$.
As already explained, each diagrams for $\Hint^{(k)}$ has $(2k-1)$ $\Delta_0$
propagators.
Hence we have $2k-1$ alternatives to make $\Hint^{(k+1)}$-diagram from one
$\Hint^{(k)}$-diagram.
As the result, the number of possible diagrams for $\Hint^{(k+1)}$ is
$(2k-3)!!\times(2k-1)=(2k-1)!!$.
Q.E.D.

Eventually, \hintn is rewritten as
$$
\eqalign{
-i \int \Hint^{(n)}dt =
\sum_{\rm possible \atop diagrams}\bcl\int
\prod_{i=1}^{n+1}dx_i^2\prod_{j=1}^{n-1}dy_j^2{ (-i\lambda)^{2n} \over (n+1)!
}\Delta_0^{2n-1} \HALF\tilde\phi^2(x_i)\bcl.
}\eqn\hintzero
$$
where $\Delta_0$'s connect $\{x_i,y_j\}$ with cubic interactions.
Each $y_j$ integration is trivial and its result is $V$.
The summation runs over all combinations of $\Delta_0$.
In fact, each combination of $\Delta_0$ gives same result.
As stated above,
$\Hint^{(n)}$ contains $2n$ vertices, $n+1$ $\tilde\phi^2$ terms, $n-1$
internal interaction points ($y_j$) and $2n-1$ $\Delta_0$ propagators.

For $n=1$,
$$
\eqalign{
&-i\int dt \Hint^{(1)} = {(-i\lambda)^2\over2}\bcl\int dx^2 dy^2 ({ 1 \over 2
}\tilde \phi^2(x))\Delta_0(x,y)({ 1 \over 2 }\tilde \phi^2(y))\bcl
}\eqn\neqone
$$
describes nothing but
the interaction in which the $k^+=0$ mode propagates between two
$\HALF\tilde\phi^2$s (See Fig.~1).
For $n=2$, $-i\int dt \Hint^{(2)}$
describes the interaction in which three $\HALF\tilde\phi^2$s are connected
with three $\Delta_0$s (Fig.~2).
Similarly, for $n=3$, $-i\int dt \Hint^{(3)}$ describes three patterns of the
interaction in which four $\HALF\tilde\phi^2$ are connected with five
$\Delta_0$'s (Fig.~3).

As already explained, if we ignore $\phi_0$, there is no $\Delta_0$ propagator
in internal lines of Feynman diagrams.
However, the lack of $\Delta_0$ is recovered by $\Hint^{(n)}$.
Namely, $\Hint^{(n)}$ supplies any diagram which is obtained by replacing some
$\tilde\Delta$ propagators to $\Delta_0$ propagators in a diagram which is
constructed from $\tilde\Delta$ only.

There is no `zero mode loop' (See Fig.~4 for example) contribution in
$\Hint^{(n)}$.
The amplitudes of Fig.~4 diverge because of multiplication of
$\delta$-functions.
Our normal ordering prescription is defined in such a way that automatically
subtracts such a divergent diagram.

\chapter{ How to Define Normal Ordering $\bcl~~\bcl$ }

For loop-level, we must consider self-contraction in $\Hint^{(n)}(n\ge1)$.
The simplest example is two-point function with one-loop(See Fig.~5a):
$$
\eqalign{
&\int dx'\int dy'\left<T\left(\phi\contract{(x)~~{1\over2!}
(-i{\lambda\over3!}}\contract{\tilde\phi^3\contract{(x'))
(-i{\lambda\over3!}}\tilde\phi^3}\contract{(y'))~~\phi}(y)\right)
\right>\cr
&={(-i\lambda)^2\over2}\int dx' dy' \tilde \Delta(x,x')(\tilde
\Delta(x',y'))^2\tilde\Delta(y',y)
}\eqn\oneloop
$$
where $\HALF$ is a symmetric factor.
There should be an amplitude which is obtained by replacing one of the internal
lines of \oneloop\ to $\Delta_0$ (See Fig.~5b), namely,
$$
\eqalign{
&(-i\lambda)^2\int dx' dy'
\tilde\Delta(x,x')\tilde\Delta(x',y')\Delta_0(x',y')\tilde\Delta(y',y).
}\eqn\onelooptwo
$$
Substituting the definition of $\tilde\Delta$ \prop to \onelooptwo, we obtain
$$
\eqalign{
\HALF{i\lambda^2\over
m^2V^2}(x^+-y^+)\sum_{p^+>0}({1\over2p^+})^3\expl{-i(p^+(x^--y^-)+{m^2\over2p^+}(x^+-y^+))}.
}\eqn\oneloopthree
$$
Note that internal propagator $\tilde\Delta(x',y')$ provides the factor $\HALF{
1 \over 2p^+ }$ in momentum space (external propagators provide the factor
${1\over 2p^+}$).
The factor $\HALF$ comes from
$$
\eqalign{
\lim_{x'{}^+\rightarrow y'{}^+}\Theta(x'{}^+-y'{}^+)=\HALF.
}\eqn\halfs
$$
The amplitude \onelooptwo\ is obtained from self-contraction of $\Hint^{(1)}$
as
$$
\eqalign{
&\left<T\left(\phi(x)(-i\int dt \Hint^{(1)})\phi(y)\right)\right>\cr
&={(-i\lambda)^2\over2}\int dx' dy'
\left<T\left(\phi\contract{(x)~~\bcl\HALF\tilde\phi^2}(\contract{x')
\Delta_0(x',y')\HALF\tilde\phi^2}(\contract{y')\bcl~~\phi}(y)\right)\right>.\cr
}\eqn\oneloopthree
$$
In order to get correct factor \onelooptwo, $\Hint^{(1)}$ should be
normal-ordered as
$$
\eqalign{
&-i\int dt\Hint^{(1)}=
{(-i\lambda)^2\over2}\int dt \bcl\sum_{p^+>0} \tilde\phi_{-p^+}\tilde\phi_{p^+}
{-i\over m^2V} \sum_{q^+>0} \tilde\phi_{-q^+}\tilde\phi_{q^+}\bcl\cr
\equiv&{(-i\lambda)^2\over2}\int dt :\sum_{p^+>0}
\tilde\phi_{-p^+}\tilde\phi_{p^+}: {-i\over m^2V} :\sum_{q^+>0}
\tilde\phi_{-q^+}\tilde\phi_{q^+}:\cr
=&{(-i\lambda)^2\over2}\int dt {-i\over m^2V} \left[:\sum_{p^+>0}
\tilde\phi_{-p^+}\tilde\phi_{p^+} \sum_{q^+>0}
\tilde\phi_{-q^+}\tilde\phi_{q^+}:+\sum_{p^+>0}:
\tilde\phi_{-p^+}{1\over2p^+}\tilde\phi_{p^+}:
\right]\cr
}\eqn\normalorder
$$
where :~~~: is the usual normal ordering, in which $\phi_p (p>0) $ should be
put on the right of $\phi_p (p<0)$.
The second term in the parenthis of the last expression of \normalorder\
contributes to \onelooptwo.
Note that, since $\Hint^{(1)}$ is instantaneous in $x^+$, an ordering of
$\tilde\phi^2(x')$ and $\tilde\phi^2(y')$ is fixed.
There is no freedom to choose $\tilde\phi^2(x')$ or $\tilde\phi^2(y')$ when we
contract with $\phi(x)$ or $\phi(y)$.
Hence we get $2\times2$ as a multiplication factor of the contraction (these
two 2 arise due to bi-linearlity of two $\tilde\phi^2$s).
In this case, naive ordering ($:\tilde\phi^2:~:\tilde\phi^2:$) gives correct
result. Unfortunately, it is not true in higher order.
The {\it normal-ordering} $\bcl~~\bcl$ of higher order interaction hamiltonians
should be decided according to the requirement of such a consistency.

Let us consider the diagram obtained from $\Hint^{(n)}$ by self-contractions.
When we contract $n_1$ $\tilde\phi^2$s in $\Hint^{(n)}$, the factor
${}_{n+1}P_{n_1}={n+1!\over(n+1-n_1)!}$ arises if the ordering of
$\tilde\phi^2$ was not fixed.
It is a desired factor, but the ordering is fixed in fact.
The factor ${}_{n+1}C_{n_1}={n+1!\over(n+1-n_1)!n_1!}$ arises
if we set
${\displaystyle\bcl\left(\sum_{p^+>0}\tilde\phi_{-p^+}\phi_{p^+}
\right)^{n+1}\bcl=
\left(:\sum_{p^+>0}\tilde\phi_{-p^+}\phi_{p^+}:\right)^{n+1}}$.
Adding to it, we have to consider the fact that
the propagator $\tilde\Delta(x,y)$ gives the factor $\HALF$
in the instantaneous limit$(x^+=y^+)$.
We must attach the factor $(\HALF)^{n_1}{}_{n+1}P_{n_1}$ to
the contracted operators {\it by hand}.
It means that
${\displaystyle\bcl\left(
\sum_{p^+>0}\tilde\phi_{-p^+}\phi_{p^+}\right)^{n+1}\bcl}$
in $\Hint^{(n)}$
must include the term
$$
\eqalign{
(\HALF)^{n_1}{}_{n+1}P_{n_1}:(\sum_{p^+>0}\tilde\phi_{-p^+}
\tilde\phi_{p^+})^{n+1-n_1}(\sum_{p^+_1>0}\tilde\phi_{-p^+_1}
({ 1 \over 2p_1^+ })^{n_1-1}\tilde\phi_{p^+_1}):.
}\eqn\termone
$$
For attaching these factors to each all contracted operators,
${\displaystyle\bcl\left(
\sum_{p^+>0}\tilde\phi_{-p^+}\phi_{p^+}\right)^{n+1}\bcl}$
is decided as follows:
$$
\eqalign{
\bcl\left(\sum_{p^+>0}\tilde\phi_{-p^+}
\tilde\phi_{p^+}\right)^{n+1}\bcl
=&:(\sum_{p^+>0}\tilde\phi_{-p^+}\tilde\phi_{p^+})^{n+1}:\cr
+&\sum_{{\rm possible}  \atop n_1,n_2,\cdots,n_I>0}
:(\sum_{p^+>0}\tilde\phi_{-p^+}\tilde\phi_{p^+})^{n+1-n_1-n_2\cdots- n_I}\cr
&\times({1\over2})^{n_1}{}_{n+1}P_{n_1}(\sum_{p^+_1>0}\tilde\phi_{-p^+_1}
({ 1 \over 2p_1^+ })^{n_1-1}\tilde\phi_{p^+_1})\cr
&\times({1\over2})^{n_2}{}_{n+1-n_1}P_{n_2}(\sum_{p^+_2>0}\tilde\phi_{-p^+_2}
({ 1 \over 2p^+_2})^{n_2-1}\tilde\phi_{p^+_2})\cr
&\vdots\cr
&\times({1\over2})^{n_I}{}_{n+1-n_1-n_2-\cdots-n_{I-1}}P_{n_I}(\sum_{p^+_I>0}
\tilde\phi_{-p_I^+}({ 1 \over 2p_I^+ })^{n_I-1}\tilde\phi_{p^+_I}):.
}\eqn\nom
$$
The factor ${\displaystyle
(\HALF)^{n_i}{}_{n+1-n_1-\cdots-n_{i-1}}P_{n_i}(\sum_{p^+_i>0}
\tilde\phi_{-p^+_i}({ 1 \over 2p^+_i })^{n_i}\tilde\phi_{p^+_i})}$
is obtained from a self-contraction of ${\displaystyle
(\sum_{p^+>0}\tilde\phi_{-p^+}\tilde\phi_{p^+})^{n_i+1}}$.

\chapter{Conclusion}

In this letter, we have considered $\phi^3$-theory as a simple example.
We show that, if we quantize the theory in a proper way,
$k^+=0$ mode propagates along internal lines of Feynman diagrams in any order
of perturbation.
The inclusion of the zero mode is expected to recover breakdown of Lorentz
invariance of a naively light-cone quantized theory after limiting procedure $V
\rightarrow \infty$.
In fact, in 1+1 dimension, the Lorentz transformation is the scale
transformation of $p^+$ and $p^-$ and it does not mix zero modes and non-zero
modes. The Lorentz invariance becomes important in higher dimensions.

The extension of the result of this paper to other theories is straightforward.
For example, in the case of $\phi^4$, we must solve the equation
$$
\eqalign{
-m^2V\phi_0 - { \lambda \over 3! }V\phi_0^3 - { \lambda\over2}\int dx^- \tilde
\phi^2 \phi_0-{ \lambda\over3!}\int dx^- \tilde \phi^3=0.
OB}\eqn\secondtwo
$$
The constraint equation becomes more complicated than $\phi^3$-theory.
\REF\Burkardt{ M.~Burkardt, Phys.~Rev. {\bf D47}(1993)4628.}
In such a case, redifinition of mass $m$ is needed( See ref\Burkardt ).

Finally we comment on the problem of vacuum (non-)triviality.
A remarkable simplicity of $\phi^3$-theory is the fact that $\phi_0$ commutes
with $\Hfree$.
In the case of $\phi^3$-theory, the time-development of $\phi$ by $\Hfree$
obeys the free equation of motion.
In $\phi^4$ (or more complicate interaction)-theory, it is not the case.
In such cases, $\Hfree$ does not reproduce correct {\it free} equation of
motion.
Hence, in general, $\Hfree$ in this letter is no more a {\it `free'  }
hamiltonian.
This discrepancy might make some clue for the problem of vacuum
(non-)triviality.
These subjects remain for future investigation.

\ack
The author would like to thank Prof.~K.~Kikkawa for valuable discussions and
careful reading of this manuscript.
He also thanks to Prof.~H.~Itoyama for useful discussions.
\refout

\vskip 3cm
\centerline{{\fourteenpoint FIGURE CAPTIONS}}
\item{Figure 1.} Two $\HALF\tilde\phi^2$ are connected by a $\Delta_0$
propagator (broken line).
\item{Figure 2.} There $\HALF\tilde\phi^2$ are connected by three $\Delta_0$
propagators.
\item{Figure 3.} Four $\HALF\tilde\phi^2$ are connected by five $\Delta_0$
propagators. Three patterns exist.
\item{Figure 4.} Examples in which $\Delta_0$ propagators make a loop by
themselves.
\item{Figure 5a.} One-loop diagram with two external lines and two internal
lines.
\item{Figure 5b.} One of internal lines of Figure 5a is replaced to $\Delta_0$.
\end